\documentclass{egpubl}
\usepackage{eurovis2024}

\ConferencePaper        
\SpecialIssuePaper         

\CGFccby

\usepackage[T1]{fontenc}
\usepackage{dfadobe}  
\usepackage{cite}  
\UseRawInputEncoding
\BibtexOrBiblatex
\electronicVersion
\PrintedOrElectronic
\ifpdf \usepackage[pdftex]{graphicx} \pdfcompresslevel=9
\else \usepackage[dvips]{graphicx} \fi

\usepackage{egweblnk}
\usepackage{dirtytalk}
\usepackage{mdframed}
\usepackage[pdftex,dvipsnames]{xcolor}  
\usepackage[most]{tcolorbox}

\usepackage{enumitem}
\usepackage{soul}
\usepackage[export]{adjustbox}
\usepackage{cancel}
\usepackage [english]{babel}
\usepackage [autostyle, english = american]{csquotes}
\MakeOuterQuote{"}

\usepackage{egweblnk}

\usepackage{microtype}
\usepackage[most]{tcolorbox}
\usepackage{pifont}
\usepackage{wrapfig}
\usepackage{tabularx}
\usepackage{hhline}
\usepackage{wrapfig}

\usepackage[T1]{fontenc}
\usepackage{dfadobe}  
\usepackage{fontawesome}

\usepackage{hyperref}
\usepackage{blindtext}
\newcommand{\dimension}[5]{
    \begin{tcolorbox}[skin=bicolor,fonttitle=\bfseries,coltitle=black,colbacktitle=#2,colback=#2!20,colframe=#2,title=#1, after skip=0.35em,left=3pt, right=3pt, top=3pt, bottom=3pt,boxsep=0pt,colbacklower=#2!10,middle=0.4em,toptitle=6pt, bottomtitle=4pt,sharp corners=all,boxrule=0mm, leftrule=1mm]
    \emph{Question}: #3
    \\
    \emph{Values}: #4
    \end{tcolorbox}%
    \noindent#5
}

\usepackage{tikz}
\usepackage{tcolorbox}
\usepackage{orcidlink}

\usepackage{subcaption}
\definecolor{agency}{HTML}{d95f02}
\definecolor{interaction}{HTML}{7570b3}
\definecolor{adaptation}{HTML}{1b9e77}

\usepackage{lipsum}                     
\usepackage{xargs}                      
\usepackage[colorinlistoftodos,prependcaption,textsize=tiny]{todonotes}

\definecolor{checked}{HTML}{22AA22}
\newcommand{\change}[1]{\textcolor{black}{#1}}
\usepackage[most]{tcolorbox}

\usepackage{framed}

\renewenvironment{leftbar}[1][\hsize]
{
    
    \MakeFramed{\hsize#1\advance\hsize-\width\FrameRestore}
}
{\endMakeFramed}
\usepackage{ragged2e}

\usepackage{mdframed}
\usepackage[most]{tcolorbox}
\makeatletter
\newtcbox{\kwColorBox}[1][]{on line,fontupper=\footnotesize\sffamily\bfseries\small,boxrule=0.5pt,arc=2pt,coltext=#1,colback=#1!10!white,colframe=#1,boxsep=0pt,left=1.5pt,right=1.5pt,top=1.5pt,bottom=1.5pt}
\makeatother

\newcommand{\kw}[2]{%
    \begin{kwColorBox}[#2]%
    {#1}%
    \end{kwColorBox}%
    \xspace%
}

\definecolor{ColorUser}{HTML}{A42C2C}
\definecolor{ColorChallenge}{HTML}{A02FA5}
\definecolor{ColorLmChallenge}{HTML}{6B4E90}
\definecolor{ColorTask}{HTML}{408E2F}
\definecolor{ColorLmTask}{HTML}{AA7A39}
\definecolor{ColorWidget}{HTML}{4F85C3}

\newcommand{\rawuserdonotuse}[1]{\kw{#1}{gray}}
\newcommand{\defLFW}[1]{\rawuserdonotuse{\phantomsection\label{user:#1}#1}}
\newcommand{\refLFW}[1]{\hyperref[user:#1]{\rawuserdonotuse{#1}}}

\newcommand{\rawchallengedonotuse}[1]{\kw{#1}{gray}}
\newcommand{\defRO}[1]{\rawchallengedonotuse{\phantomsection\label{challenge:#1}#1}}
\newcommand{\refRO}[1]{\hyperref[challenge:#1]{\rawchallengedonotuse{#1}}}

\title[Deconstructing Human-AI Collaboration]%
      {Deconstructing Human-AI Collaboration: \\ Agency, Interaction, and Adaptation \\ }

\author[S. Holter \& M. El-Assady]
{\parbox{\textwidth}{\centering Steffen Holter~\orcidlink{0009-0008-2935-5549}
        and Mennatallah El-Assady~\orcidlink{0000-0001-8526-2613} 
        }
        \vspace{-0.5em}
        \\
{\parbox{\textwidth}{\centering
         ETH Zurich
       }
}
}

\begin{document}


\teaser{
\vspace{-3.5em}
  \center
  \includegraphics[width=1\linewidth]{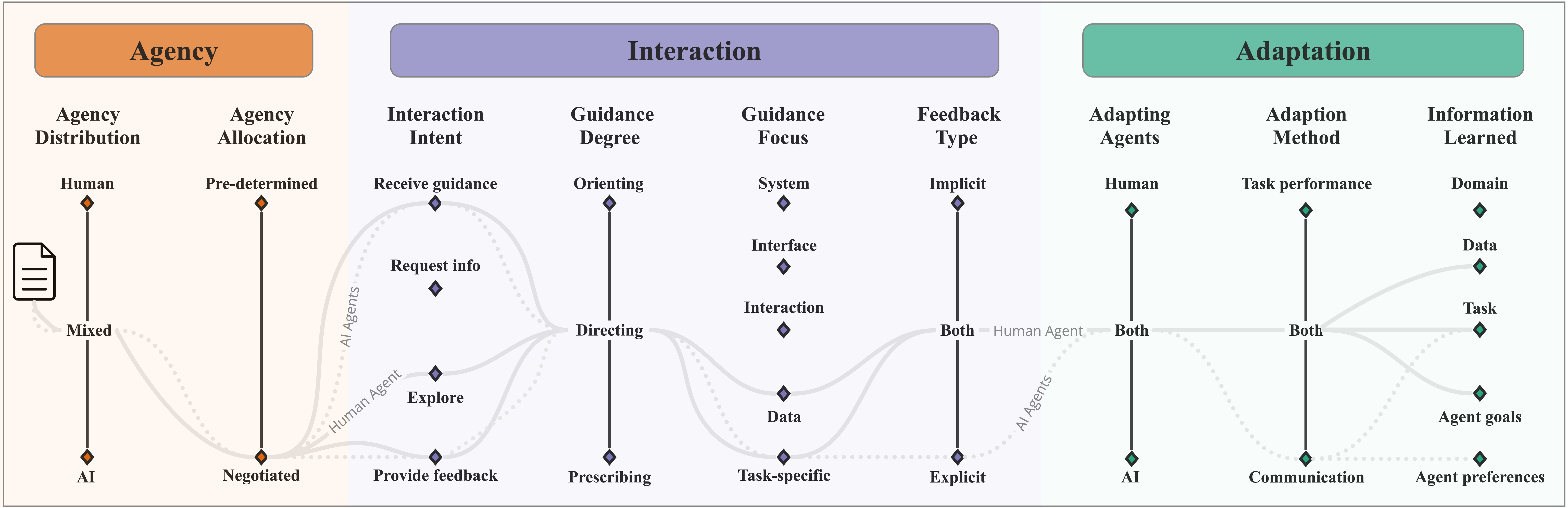}
  \vspace{0.000001pt}
  \caption{\textbf{Conceptual Model}: The design space of human-AI collaboration is comprised of three high-level \change{categories}: (1)~agency, (2)~interaction, and (3)~adaptation, which are in turn split into \change{specific individual} dimensions. Each vertical dimension is described by either a spectrum or a set of categories and reflects a \change{distinct} property of human-AI collaboration. An exemplary mixed-initiative system for \textbf{topic model refinement} is encoded using parallel coordinates to illustrate a typical use case of our design space. The conceptual model is described in detail in \autoref{sec:designspace}.}
  \vspace{0.3em}
  \label{fig:conceptual_model}
}

\maketitle

\begin{abstract}
As full AI-based automation remains out of reach in most real-world applications, the focus has instead shifted to leveraging the strengths of both human and AI agents, creating effective collaborative systems.
The rapid advances in this area have yielded increasingly more complex systems and frameworks, while the nuance of their characterization has gotten more vague.
Similarly, the existing conceptual models no longer capture the elaborate processes of these systems nor describe the entire scope of their collaboration paradigms.
In this paper, we propose a new unified set of dimensions through which to analyze and describe human-AI systems.
Our conceptual model is centered around three high-level aspects - agency, interaction, and adaptation - and is developed through a multi-step process.
Firstly, an initial design space is proposed by surveying the literature and consolidating existing definitions and conceptual frameworks.
Secondly, this model is iteratively refined and validated by conducting semi-structured interviews with nine researchers in this field.
Lastly, to illustrate the applicability of our design space, we utilize it to provide a structured description of selected human-AI systems.




\end{abstract} 



 \section{Introduction}
Collaboration between humans and AI is as rich with opportunity as it is with ambiguity. 
It is widely agreed that by exploiting the complementary capabilities of humans and AI, it is possible to achieve joint performance superior to that of fully manual or completely automatic systems. 
While several frameworks ~\cite{reverberi_experimental_2022, sowa_cobots_2021, lee_human-ai_2021} have been built bearing the human-AI collaborative label, there is currently no systematic way of comparing and classifying these systems. 
Generic terms such as human-AI teaming \cite{xu_applying_2023}, hybrid intelligence~\cite{dellermann_hybrid_2019}, and collaborative decision-making~\cite{schemmer_towards_2023} are often used interchangeably in literature while the characteristics of these systems do not necessarily align. 
Conversely, the plethora of analogous terms creates parallel research communities that are discussing equivalent concepts in isolated settings. 
This lack of coherence makes it challenging to examine and review the field in any meaningful way. 
Similarly, the existing conceptual models fall short of incorporating the full scope of systems due to this ambiguity in both definition and terminology. 


In this paper, we introduce a novel conceptual model to structure the design space of human-AI collaboration, providing a systematic approach for describing and discussing such systems. 
We show that by deconstructing this complex interaction paradigm first into three higher-level categories---agency, interaction, and adaptation---and then into specific sub-dimensions, we provide a comprehensive means of encoding all human-AI systems. 
A \change{complete} overview of our conceptual model is shown in \autoref{fig:conceptual_model}.
Our proposed design space caters to researchers and experts in the field looking to reason about their systems, design choices, or conceptual frameworks. 


Our methodology to derive the design space looks to combine existing conceptual frameworks in this research area to find a meaningful intersection of definitions and generate an initial prototype model. 
Building on top of a systematic literature review, we iteratively refine our set of dimensions by conducting interviews and receiving feedback and comments from \change{expert} practitioners in the field. 
Thereby, we look to establish a consensus that also reflects the opinions and approaches of the extended research community working in human-AI collaboration.

Since the motivation for \change{a unified design space} stems from the need to describe the characteristics of human-AI collaboration, we evaluate our approach by analyzing a selection of popular frameworks. 
The classification of existing systems within the scope of our proposed design space also acts to validate the specific choice of dimensions. 
If certain proposed dimensions are perpetually latent, then it suggests that either these characteristics are not relevant or feasible in practical application or that they present a potentially interesting research opportunity that can be tackled in the future. 
Thus, in addition to refining the research space for human-AI collaboration, we are scoping out potential challenges, research gaps, and future directions within the field. 

    


Overall, our main contributions are as follows: (1)~\textbf{A conceptual model for human-AI collaboration}, which models the design space of human-AI collaboration. (2)~\textbf{A two-step \change{methodology}} consisting of (a) collection and analysis of existing human-AI collaboration literature for building the initial \change{prototype design space}, and (b)~\change{iterative} refinements by incorporating feedback from researchers in the field. (3)~\textbf{Case studies} using the proposed conceptual model to analyze three human-AI collaborative systems.



In sum, we provide a solution that unifies the diverse design considerations in the field of human-AI collaboration under a single comprehensive conceptual model. 
We aspire to meaningfully contribute to the community's capacity to discuss, design, and analyze present and future human-AI systems. 

 \section{Related Works}

While no unified body of work addresses the entire human-AI collaboration paradigm, a lot has been done to analyze various key components of this collaboration.

\noindent\textbf{Human-AI Interaction ~|~} A general set of design guidelines concerning how human-AI interaction should be implemented is proposed and evaluated by Amershi et al.~\cite{amershi_guidelines_2019}. This set of principles provides a base from which practitioners can build human-facing systems. \change{There is also work focused on understanding how end-users engage with predictive systems~\cite{lu_state_2017}}. Others focused on articulating why designing these interaction paradigms is so difficult ~\cite{yang_re-examining_2020} and what actually is expected from AI collaborators in various domains (e.g., medicine) ~\cite{cai_hello_2019, maadi_review_2021, ashktorab_effects_2021}. El-Assady et al.~\cite{el2022biases} explore the details of the communication processes and propose dynamic models of explanation intent and receiver perception. In an attempt to model this complex interaction, researchers have even looked towards human-human communication for inspiration ~\cite{wang_human-human_2020, mou_media_2017}. Decision-making through interaction has also been surveyed from the perspectives of tasks, AI elements, and evaluation metrics ~\cite{lai_towards_2023}.

\noindent\textbf{Adaptive Systems ~|~}  To effectively improve collaboration requires adaptation and learning from both the \change{side of the} human and AI. The process also referred to as co-adapting, has been looked at from \change{multiple} perspectives. For example, Sperrle et al.~\cite{sperrle_co-adaptive_2021} \change{aimed} to model this type of combined learning by deconstructing it into incremental learning goals. Since \change{adaptation} often happens in response to human feedback, it is useful to analyze how this happens and what are the contributing factors ~\cite{metz_rlhf-blender_2023}. Other research directions include focusing on understanding what the AI actually has learned~\cite{cabrera_what_2023} and how knowledge is even generated ~\cite{sacha_knowledge_2014}. In practical application, this adaptation is already widespread and forms the basis of, for example, active learning systems~\cite{bernard_comparing_2018}.

\noindent\textbf{Mixed-Initiative Systems ~|~}  Horvitz’s formative paper on mixed-initiative systems ~\cite{horvitz_principles_1999} proposed principles for distributing agency within systems. \change{These} mixed-initiative approaches have become ubiquitous, especially in visual analytics (VA). Such systems have been built to rank data~\cite{wall_podium_2018}, check facts ~\cite{nguyen_believe_2018}, translate text ~\cite{coppers_intellingo_2018} and improve models ~\cite{gou_vatld_2021, ming_protosteer_2020}, to just name a few. Others have contributed by creating structured evaluation frameworks for these VA systems as a basis for comparison and analysis ~\cite{sperrle_survey_2021}. Monadjemi et al.~\cite{monadjemi_human-computer_2023} look at collaboration in VA through an agent-based lens, which has shown to yield itself well to more complex mixed-initiative and guided collaborative systems. \change{While Wang et al.~\cite{wang_survey_2022} take a more general approach and survey different ways ML agents can be integrated into VA tasks.} The recent prominence of mixed-initiative systems has inspired a new corpus of work relating to guidance --- supporting user's analysis tasks --- in these systems ~\cite{sperrle_learning_2020, perez-messina_typology_2022, perez-messina_methodology_2023}. This supplements the already considerable body of work describing guidance in more traditional VA systems~\cite{ceneda_characterizing_2017, ceneda_review_2019, sperrle_lotse_2022}. All of these insights inform the concept of initiative in more general human-AI collaborative systems. Agency in a broader context has yet to be explored in-depth but the trade-off between balancing control and autonomy has already been chronicled in some human-AI systems ~\cite{li2023we}.

\begin{figure*}[h]
  \centering
  \includegraphics[width=0.98\linewidth]{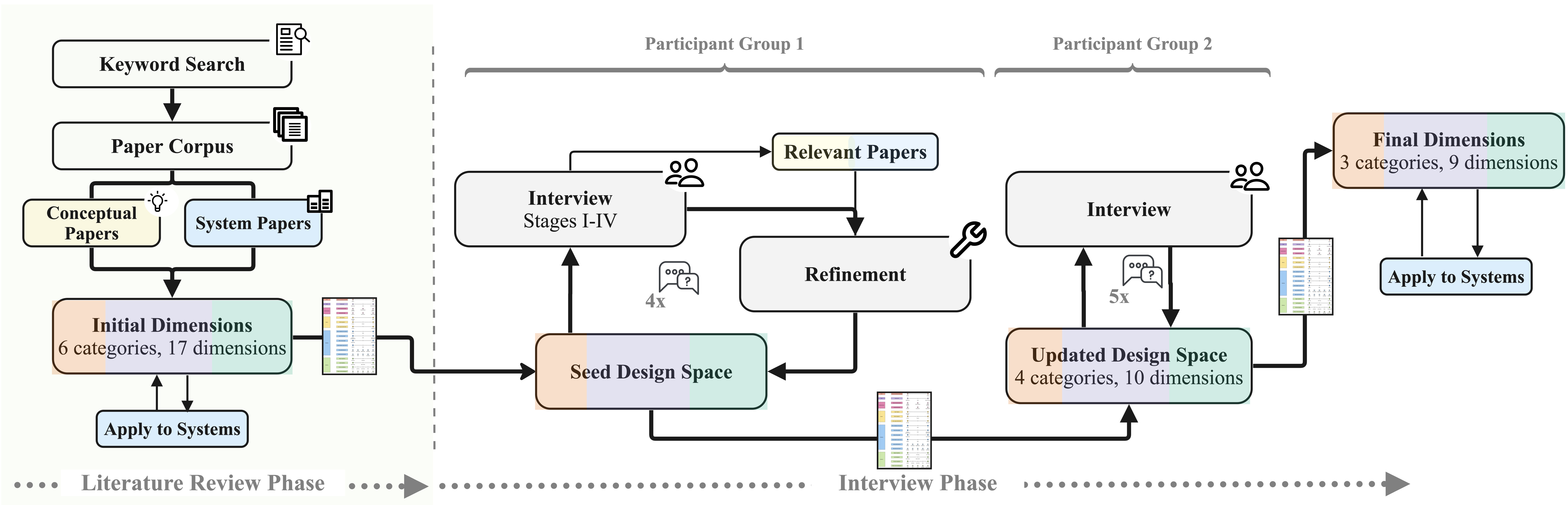}
  \vspace{.5em}
  \caption{\textbf{Methodology Overview}: A two-phase process to formulate a design space for human-AI collaboration. The \textbf{Literature Review Phase} consists of assembling a corpus of relevant papers and extracting an initial set of dimensions. The \textbf{Interview Phase} is used to iteratively refine our design space through collaboration. The interviews were conducted in two participant groups: (a) where modifications and improvements were applied after each iteration and (b) where feedback was aggregated over the course of multiple interviews. The entire process converged at a compact final set of dimensions that make up our design space.} 
  \vspace{-.5em}
  \label{fig:methodology}
\end{figure*}

 \section{Methodology}

The process for constructing our conceptual model to represent the design space of human-AI collaboration is split into two distinct phases, broadly outlined in~\autoref{fig:methodology} and described in detail below. 
First, we survey relevant literature to map out the current trends, definitions, and conceptual frameworks within this research space. 
Using this, we aggregate the findings into an initial set of dimensions that could be used to describe this collaboration paradigm. 
Second, we conduct interviews with researchers in the field of human-AI interaction to discuss our initial prototype and iteratively refine it based on their feedback and comments. 
In parallel, the completeness of each iteration of the design space is evaluated by utilizing it to characterize select human-AI systems. 
We use this two-phase approach to incorporate input from the community and to establish a conceptual model that could be generalized to the requirements of the field as a whole. 
When developing our design space, a key challenge is establishing the proper granularity of our dimensions. 
It is paramount that we provide enough flexibility to distinguish between different systems while maintaining enough generalizability to find commonality between similar frameworks existing in this space. 
For example, it is useful to consider the notion of \textit{guidance} as being relevant to human-AI systems. 
However, completely deconstructing guidance from the perspective of processes or visual analytics into sub-categories can be deemed excessive and even counter-productive in our efforts to \change{create a unified} design space. 
Thus, each choice of dimension represents a trade-off between expressiveness and complexity for our conceptual model.

 \subsection{Systematic Literature Review}

In this first phase, we performed a general keyword search using terms like "\textit{human-AI}, \textit{human-AI collaboration}, \textit{human-AI teaming}, \textit{collaborative decision-making}, \textit{hybrid intelligence}, \textit{human-AI interaction}, \textit{mixed-initiative}, and \textit{collaborative analytics}" on the titles, abstracts, as well as contents of recently published works (since 2010), retrieving an initial set of potentially relevant papers.
We supplemented this set by manually parsing the proceedings of relevant and widely recognized conferences, including CHI, IUI (TIIS), EuroVis (CGF), and VIS (TVCG), to collect works that do not align with the terminology used in our search but were deemed as pertaining to human-AI collaboration. 
Then, by manually reviewing and selecting papers that fit into our broad definition of human-AI collaboration, we compiled an initial corpus of relevant works. Additionally, relevant referenced papers from the original seed set were inspected and included if \change{deemed important}. 
The assembled collection was broadly categorized into two sets of papers: (1)~conceptual frameworks and models and (2)~human-AI systems and techniques. The conceptual papers include works that address this research space through surveys, taxonomies, or other high-level approaches. The rest of the papers represent a range of examples of collaborative systems and techniques that we used to validate our proposed design space.

The extraction of a concrete set of dimensions from the assembled list of papers was done by first \change{manually} aggregating all the different \change{keywords}, properties, and descriptions mentioned in the set of conceptual papers. 
Due to ambiguities in the definition, the initial collection featured a lot of redundancy and was manually corrected by collapsing similar dimensions together. 
Another round of refinements focused on isolating the properties that describe the human-AI systems themselves and not the motivations or underlying context behind the design decisions. 
\change{For instance, the type of end user (e.g., domain expert, novice) motivates certain system design choices (e.g. the degree of guidance the system provides) but does not describe the system itself and thus is not an appropriate dimension in our conceptual model.}
This process yielded an assortment of 15-20 properties \change{that were} clustered into five higher-level \change{categories} for easier \change{evaluation}. 
As a sanity check, this prototype design space was applied to select systems from the second pool of papers collected, and working through the process allowed us to prune our conceptual model further. 
At the end of this check, we were left with 17 dimensions assembled into five categories, which \change{acted as the seed design space for the interview study} and can be seen in its entirety in Appendix A.


 \subsection{Interview Study}

The second phase involved semi-structured interviews with practitioners and researchers in related fields. 
Here, we describe the properties of the study setup, the participants, the various stages of the study, and our iterative method. 

\noindent\textbf{Participants ~|~}  In total, the interview study consisted of discussions with nine PhDs and Post-Docs. 
At the start of the interview, we asked the subjects to self-identify their specific field(s) of expertise to have a high-level overview of what domains are primarily informing their thinking. 
The majority featured backgrounds in visual analytics, explainable AI, and NLP. However, some also considered themselves specialists in reinforcement learning, GNNs, and general machine learning. 
The resulting distribution covers a wide spectrum of fields that directly inform human-AI collaboration or lie adjacent to it. 

\noindent\textbf{Study Setup ~|~} 
The interviews took place both in-person and online, lasting approximately 1-1.5 hours each. 
To provide a uniform structure, the entire process was carried out using a \textit{Miro} collaborative sketch board. 
During the discussion, the participants were asked to continuously summarize their main ideas using post-its while vocalizing their general thought process to the interviewer. 
This helped us to distill the key arguments as participants were forced to formalize their complex ideas in very specific terms. 

\noindent\textbf{Interview Stages ~|~}  To reaffirm the semi-structured nature of the study, the entire process was split into four concrete stages (I-IV). Each part corresponded to a specific higher-level goal and was also characterized by the amount of input provided by the interviewer. The complete breakdown of this process is described below.

\begin{enumerate}[wide]
    \item[I.~\textbf{Introduction}]- This preparatory stage gave a general introduction to the interview structure, the aims, and the motivation for the project.
    Specifying that the goal is to establish and refine a conceptual model allowed the participants to start thinking in terms of so-called dimensions right from the start. 
    For those who were not familiar with the Miro board framework, we described the mechanisms and filled in some metadata information (i.e., name, title, etc).
    \item[II.~\textbf{Ideation}]- Next was a brainstorming phase where we asked the participants to describe some human-AI collaborative systems they are familiar with. 
    This information gives valuable insight into what people consider to fall under this generic term. 
    Then, we asked participants to deconstruct these systems into a set of properties or keywords. 
    At this phase of the interview, the facilitator's guidance level was kept to a minimum to allow for unbiased ideation. 
    When designing this process, we acknowledged the difficulty of eliciting novelty in real-time, and thus, we expected that participants would only be able to address the human-AI interaction paradigm partially.
    \item[III.~\textbf{Thinking in Dimensions}]- After compiling the first iteration of ideas from the interviewees, we started to guide them toward constructing more explicit dimensions. 
    This entailed progressing from specific system properties to larger groupings of potential sets of characteristics. 
    We revealed our high-level categories and an example dimension - \textit{Agency Distribution} to further assist with this process. 
    The participants could choose whether to align their thinking with \change{our example categories} or propose their original groupings. 
    At this stage, the interviewer became more active in the discussion and helped steer the subjects to keep thinking in terms of dimensions.
    \item[IV.~\textbf{Design Space}]- The first three stages acted to extract an unbiased collection of insights and ideas. 
    It also allowed the participants to formulate their \change{own} mental model of what human-AI collaboration entails. 
    Conversely, the final stage consisted of presenting our proposed design space and examining how it compares to the dimensions suggested by the participants. 
    To provide coherency, each high-level category was first described in its entirety and then reviewed together. 
    All the feedback was noted down, but to foster a meaningful discussion, the interviewer also presented rebuttals to the suggestions and criticisms when appropriate. 
    The subjects were also encouraged to add, manipulate, and re-arrange the dimensions.
    
\end{enumerate}

\noindent\textbf{Iterative Approach ~|~}  The full interview study was used to iteratively refine and improve the initial design space generated in the first phase. 
The total pool of participants was randomly separated into two groups. 
For the first batch of 4 people, we refined the conceptual model after each interview based on feedback and comments. 
Since the entire design space was less polished in the initial phases, this iterative approach allowed us to rapidly converge to a more convincing set of dimensions. 
The second batch of 5 people were all presented with the same improved design space, and the suggestions were simply collected and addressed in aggregated form after the completion of the study. 
Discussing the same conceptual model with different subjects acted to consolidate our proposal and stress test it against an assortment of opinions.
\change{The complete evolution of the design space and the high-level categories throughout the interview study is depicted in Appendix B.}
As a complementary part of the refinement process, the design space was again applied to the same selection of human-AI frameworks after each iteration to ensure that it retained sufficient flexibility to outline these systems.

\section{Design Space of Human-AI Collaboration}\label{sec:designspace}

In this section, we outline our proposed conceptual model. 
Our solution is built around three fundamental high-level aspects of human-AI collaboration - agency, interaction, and adaptation. 
Through the iterative two-phase \change{refinement} process, we established that these domains encompass the fundamental characteristics needed to describe most human-AI systems. 
To provide our framework with sufficient flexibility, these categories are then further broken down into a set of dimensions, with each signifying specific system properties. 
The complete design space with an example topic refinement system ~\cite{sperrle_learning_2021} is summarized in~\autoref{fig:conceptual_model}. 

Each individual dimension incorporates either a continuous range of values or a categorical itemized list. 
However, while some characteristics exist on a spectrum, for this design space, we simplify the continuous range to certain fixed values. 
The minimum granularity that still exhaustively describes each property is selected. 
This makes it easier to systematically quantify the position of a particular system on the spectrum. 

The described design space is not completely orthogonal. 
This is not a limitation of our model but rather a reflection of the complexity associated with human-AI collaboration. 
As such, depending on the system being analyzed, some dimensions will not be required due to the absence of certain system characteristics. 
For example, a system that lacks adapting agents will have latent dimensions in the adaptation section of the design space.  

\begin{wrapfigure}{r}{0.25\textwidth}
    \vspace{-13pt}
    \hspace{-45pt}
    \centering
    \includegraphics[width=1\linewidth]{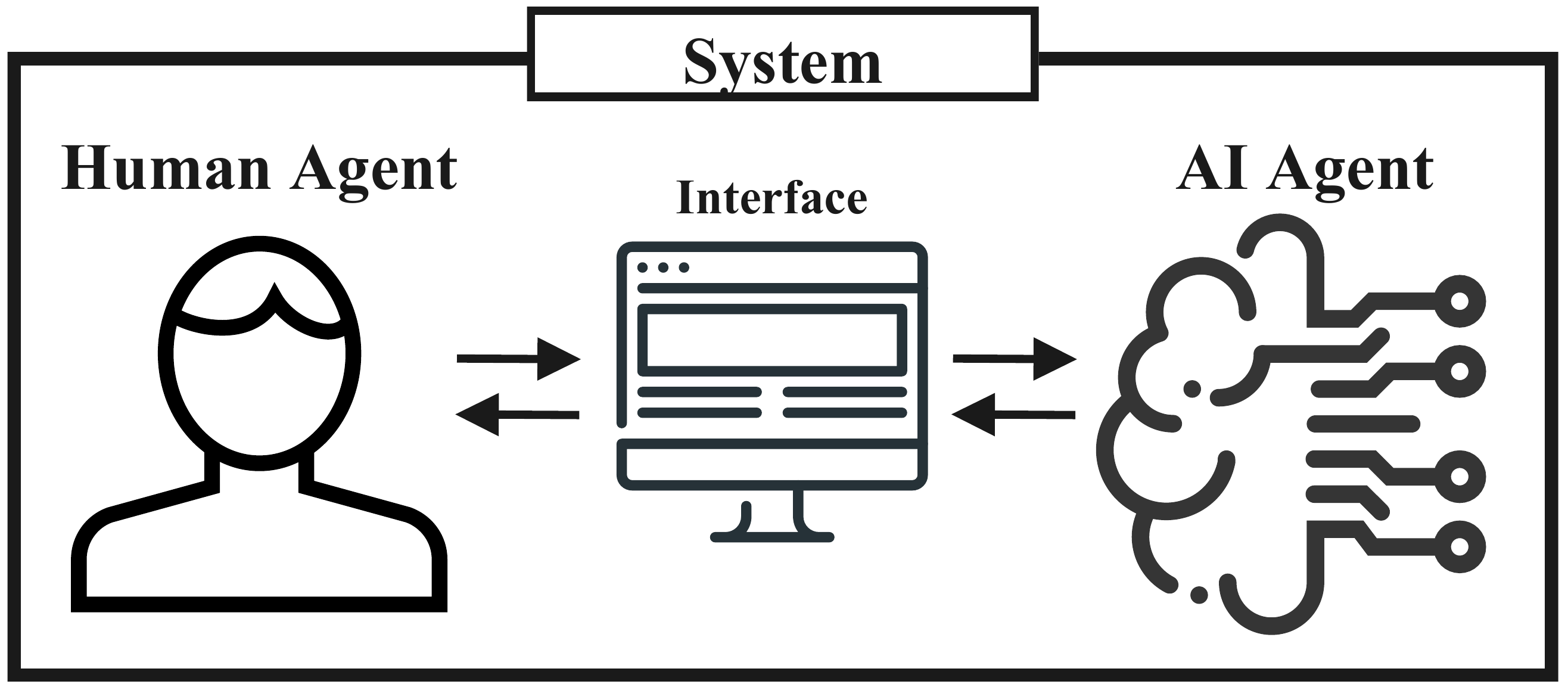} 
    \vspace{-10pt}
    \hspace{-25pt}
    \label{fig:sys}
\end{wrapfigure}

Before detailing the conceptual model, it is worth clarifying some terminology and our general framing of human-AI systems. 
By default, all agents are considered equal participants in the system communicating via some type of interface (e.g., visual user interface). 
Every agent participating in the collaboration is modeled in parallel, and the aggregation of these individual agents acts to describe the system as a complete whole. 
The benefit of this approach is that it provides a means to model systems on a per-interaction level while also establishing a maximum capacity for the entire system.
We do distinguish between two general types of agents, namely human and AI; however, to ensure symmetry in our design space, no attributes are assigned based simply on agent type. 
Furthermore, our parallel formulation allows us to efficiently accommodate any number of different agents \change{in multi-agent systems}.

The following sub-sections provide an in-depth analysis of each of the three high-level categories. First, we highlight how certain works and concepts from the literature review stage have directly informed the choice of dimensions. 
Next, each dimension is reasoned about individually and broken down into questions and sets of values. 
Any ambiguity is reduced by explicitly defining the component terminology. 
Finally, we summarize the interview feedback and observations regarding each set of dimensions.
This part is structured into specific themes that were extracted by manually aggregating the participants' comments and questions. 
Additionally, throughout the presentation of our design space, we look to highlight potential research opportunities \defRO{RO}, and limitations \& future work \defLFW{LFW}.





\subsection{Agency}

At the highest level, the term agency refers to who is in control during the analysis process (i.e., decides how to proceed) and has the responsibility of decision-making. Many human-AI systems that are used for applications such as model debugging ~\cite{hohman_gamut_2019, ming_rulematrix_2019,wexler_what-if_2019} remain human-centered without automated components. However, an emerging sub-group of human-AI collaborative systems called mixed-initiative systems has looked to give AI agents more responsibility to reduce the human task load. For example, Sperrle et al.~\cite{sperrle_co-adaptive_2021} proposed a system where multiple AI agents are assisting in a topic model refinement task while Podium~\cite{wall_podium_2018}, represents a system for ranking data points based on their perceived value.
Agency has also been considered from a psychological perspective and the tension that exists between users' desire to automate certain actions and the hesitation to cede decision-making control~\cite{sundar2020rise}. This trade-off is also mentioned in the descriptions of some human-AI systems ~\cite{li2023we}. 
While agency is indirectly mentioned in many system descriptions, there is a lack of work addressing this topic specifically and looking at the trade-off that exists between agency and automation \refRO{RO1}.  


\dimension{Agency Distribution}{agency}{Who has the agency in the system?}
{Human, AI, Mixed}

This dimension establishes the general distribution of agency within any given system based on \change{whether} the agents are allowed to make decisions. 
While the \textit{human} and \textit{AI} options represent the fully manual and fully automated edges of the agency spectrum, the \textit{mixed} category encompasses all situations where the system allows for both types of agents to assume control at some point during the interaction. 
Within human-AI collaboration paradigms, the AI's role has often been reduced to the application of its computational ability (e.g., computation, search, storage, etc.) to assist human users in solving analytical tasks. 
However, recently, there has been a shift where AI has been given autonomy about simpler, repetitive, and taxing tasks in an attempt to alleviate the burden on humans. 
In other words, the AI is no longer treated as a \textit{tool} but rather as a participating \textit{agent} who reasons about the analytic process and takes actions \change{in tandem with} human users. This motivates the need for this dimension in order to effectively distinguish between modern systems within this research space. 


\dimension{Agency Allocation}{agency}{How is agency determined within the system?}
{Pre-determined, Negotiated}

The choice of \textit{who} is in charge only informs the general organization of the system. 
Yet, in the cases where the framework features a mixed-initiative approach, an additional dimension is required to describe how the agency is decided over the course of the collaboration. 
\textit{Pre-determined} characterizes systems where the orchestrator is designed to assign agency according to a static set of established criteria. 
For example, for a semi-autonomous vehicle, the human might drive in the city and the AI on the highway. 
This set of rules is usually already incorporated into the system by the developers of the system itself.
On the other hand, a \textit{negotiated} approach refers to a dynamic way of establishing agency where the agents’ roles are not determined in advance but opportunistically decided while solving the task.
While such an approach theoretically introduces flexibility into the system, it also presents several challenges for practical implementation. 
For instance, the conditions of the negotiation still need to be formulated in advance by the system designers. 
Ideally, for a symmetric collaboration process, the decision for which agent is best suited at any given time should be based on a quantifiable objective function. 
However, since most analytical tasks involve considerable ambiguity and multiple conflicting objectives, the choice of agency can become subjective and a matter of preference. 
Additionally, not only should this negotiation consider who is best suited, but also what else could they be doing instead (i.e., opportunity cost).
Thus, building a systematic way of carrying out the opportunistic negotiation of agency presents a research opportunity \refRO{RO1} in mixed-initiative system design.


\subsubsection*{Participant Comments and Interview Observations} 
\begin{leftbar}
    The general consensus was an agreement with our proposed dimensions, yet during the ideation stages (II \& III), some participants also proposed additional questions. The two dimensions of \textit{who is deciding agency?} and \textit{when is agency decided?} were suggested most frequently by the practitioners. 

\textbf{Assigning agency ~|~}  In our dimensions, we are not specifying who is allocating the agency during the interaction. However, during the discussions with practitioners, it became apparent that in the majority of cases, this orchestrator role is simply fulfilled by the system itself. Thus, while it is not distinctly stated, it can be considered implicitly implied in our design space.

\textbf{Temporal dimension ~|~}  The matter of \textit{when} was something briefly examined in the prototyping phase. Specifically, the benefit of considering the temporal aspects is that it gives a more complete picture of the distribution of agency. For example, in many systems, the agent who is in control depends on what phase of the interaction they are in. Eventually, the dimension was omitted due to high variance and a very strong dependence on the specific analytic task and domain. Furthermore, it can be argued that many elements of \textit{when} are contained within our second dimension as designing negotiation or pre-determining agency incorporates some temporal considerations. Regardless, constructing a structured temporal axis for the entire conceptual model is an opportunity worth trialing in the future \refRO{RO2}.
\end{leftbar}



\subsection{Interaction} 

The interaction category examines the specific ways that human and AI agents communicate and collaborate with each other to solve analytical tasks. 
There is considerable complexity embedded within this notion and modeling it requires an interplay of different disciplines, such as machine learning, user interface design, and even psychology ~\cite{dudley_review_2018}. 
Furthermore, to reduce ambiguity, we attempt to establish some boundaries for what we are considering as interaction. Instead of modeling an entire sequence of complex interactions, we focus on smaller sets of lower-level interaction events that characterize a system's general operation. While these events are semantically poor, they can be captured, grouped, and reasoned about more effectively. This granularity of interaction is inspired by Gotz \& Zhou's approach for visual analytics ~\cite{gotz_characterizing_2008}. Similarly, previous work in VA has looked at not only categorizing low-level interactions but also identifying users’ motivations~\cite{brehmer_multi-level_2013}. Another approach has looked at deconstructing interaction logs and establishing concrete start and end points of interactions~\cite{gathani_grammar_2022}. 

To understand interaction, it is fundamental also to address the notion of communication. Interdisciplinary research has looked at how human-AI communication can learn from human-human interaction ~\cite{wang_human-human_2020}. In particular, El-Assady et al.~\cite{el2022biases} proposed a communication dynamics model that makes a clear distinction between perspectives (i.e., sender or receiver). This framing of communication is also incorporated into the design of our initial design space. Another key component of interaction is guidance, which at a high level can be seen as a means of closing the knowledge gap between agents. Quantifying guidance has been a feature in a number of foundational papers~\cite{ceneda_characterizing_2017} and also directly motivates our design space. Recent work has re-defined guidance as an active process that accounts for previous actions ~\cite{sperrle_learning_2020}. This also evokes the need to reason about the specific components and characteristics of feedback ~\cite{metz_rlhf-blender_2023}. 




\dimension{Interaction Intent}{interaction}{What are the interaction motives of agents?}
{Receive Guidance, Request Info, Explore, Provide Feedback}

This dimension captures all the high-level reasons agents can have for interacting within the system. An agent can \textit{receive guidance}, which we define as a process whereby the knowledge gap is resolved through gaining new information from another agent. This usually pertains to the analytical task and is either provided automatically or upon request. Thus, the second motivation for interacting can be seen as querying or \textit{requesting information} from other agents. For example, after a human user receives a suggestion regarding an analytic task, they might ask the AI to provide additional context through explanations. While the aforementioned options are ways for agents to refine their existing mental models, the cause for interaction can also be to generate hypotheses through \textit{exploration}. This includes probing the interface, viewing datasets, examining decision histories, and other actions that do not facilitate the flow of information between agents. Finally, the intention can be to \textit{provide feedback} to other agents. In most cases, this is in response to guidance and is an evaluation against some implicit or explicit metric.



\dimension{Guidance Degree}{interaction}{What is the extent of guidance being received?}
{Orienting, Directing, Prescribing}

How collaborative systems aim to resolve the knowledge gap is related to the amount of guidance that is being provided by the agents. In the corresponding literature, the degree of guidance is split into three main categories: orienting, directing, and prescribing~\cite{ceneda_characterizing_2017}. \textit{Orienting} guidance refers to instances where the agent accentuates a set of possibly relevant options but does offer a comparative analysis of the suggestions. \textit{Directing} guidance adds a ranking to the provided choices and thus includes more information regarding the relevance with respect to the analysis task. Finally, \textit{prescribing} suggests a single very specific recommendation deemed optimal by the agent while alternative options are omitted. Distinguishing between the degrees of guidance is particularly useful when characterizing systems meant for different types of users. For example, for expert users, orienting guidance is sufficient use of the AI agent's strengths (e.g., processing of data, recognizing complex patterns, etc.) while novice users might require specific prescribing predictions (e.g., a machine learning solution) to solve the analytic task. 


\dimension{Guidance Focus}{interaction}{What elements does the guidance pertain to?}
{System, Interface, Interaction, Data, Task-Specific}

In addition to the degree, it is also valuable to detail what the guidance pertains to. While the collaboration in human-AI systems is usually centered around an analytical task, the guidance provided can extend beyond solving this task. Information regarding the \textit{system} as a whole can be considered an onboarding phase where insight is shed on the operation and aims of the framework as a whole. Assistance involving the \textit{interface} specifies the features and capabilities of the tool. Similarly, guidance on \textit{interaction} consists of specifying and describing the available interactions. These first three options can also be summarized as types of general guidance. \textit{Data} focused suggestions include everything related to the raw data, such as distributions and statistics. Finally, \textit{task-specific} guidance encompasses everything related to the task. Unlike the previously mentioned options, providing task-specific (and, to some extent, data-related) guidance requires knowledge about users' intents and an identifiable target task. There also exists a correlation between the focus of guidance and the semantic complexity of a given suggestion~\cite{sperrle_lotse_2022}. 




\dimension{Feedback Type}{interaction}{What types of feedback are being provided?}
{Explicit, Implicit, Both}

Finally, in response to the guidance, an agent can provide feedback, which can be expressed in two primary ways. Firstly, the agent might produce \textit{explicit} feedback, a conscious input relating to the received guidance. This can have different modalities (visual, natural language, etc.) but is characterized by intent. Conversely, \textit{implicit} feedback is often indirect and an unintentional way of reacting to guidance. Examples include recording clicks, drags, and other interactions with the interface (i.e., provenance). The specifics of both the explicit and implicit feedback are system-specific, and a universal encoding to describe all types of inputs is not feasible ~\cite{metz_rlhf-blender_2023}.

\subsubsection*{Participant Comments and Interview Observations} 
\begin{leftbar}
Due to the nuance associated with describing interaction, this section evoked a lot of discussion and was the category that went through the most refinement during the interview stage. The main themes considered were defining interaction and symmetry. 


\textbf{Defining interaction ~|~}  In the first prototype of the framework, this category was split into smaller constituents such as communication, guidance, and feedback. Guidance was attributed to the AI and feedback to the human. However, with each iteration, it became apparent that they all represent various facets of the interaction paradigm. All of these components were thus grouped under a single interaction and switched to a per-agent modeling approach that does not depend on the type of agent.  

\textbf{Symmetry ~|~}  Building on the previous theme, a big motivation for creating one unified interaction category was the notion of symmetry. The initial framing where guidance was provided only by the AI and feedback by the human is what is observed in a lot of VA and human-AI decision-making literature ~\cite{sperrle_co-adaptive_2021}. However, this implicitly implies a human-centered bias in the collaboration. This absence of symmetry was emphasized particularly by researchers with machine-learning backgrounds who did not view humans as being inherently superior. As our goal is to model the entire space of human-AI collaboration, introducing symmetry was seen as a key requirement.

\end{leftbar}

 \subsection{Adaptation}

A distinction is often made between simple-reflex AI agents (i.e., trained just once) and learning AI agents (continuously updated to be adaptive) ~\cite{kuhl_artificial_2022}. Learning for these AI agents might incorporate updating parameters or even retraining the model.  In parallel, the human collaborators are also learning when solving the task by building unique knowledge ~\cite{schemmer_towards_2023}. Therefore, in a collaborative process, both the human and AI agents can learn from each other through various mechanisms, such as labeling, demonstrating, teaching adversarial moves, criticizing, and rewarding \cite{dellermann_hybrid_2019}.

To discuss learning goals and improving communication between agents, it is also useful to look at the relationships between goals and tasks ~\cite{lam_bridging_2018, bors_provenance_2019}. Through adaptation, human users and AI systems adapt over time to converge to a common understanding and shared analysis process to solve tasks. The specific information transferred between the agents directly informs adaptation. Modern AI systems are generating new knowledge in complex domains and thus can be as beneficial to human learning as humans are for AI learning ~\cite{dellermann_hybrid_2019}. The details of how information moves are directly dependent on how the feedback is modeled ~\cite{metz_rlhf-blender_2023}.

\dimension{Adapting Agents}{adaptation}{Who is adapting in the system?}
{Human, AI, Both}

This dimension determines which types of agents have the ability to learn during the collaboration. 
The absence of a \textit{none} option relates to the assumption that a human agent will always elicit some sort of learning when interacting with a system and solving an analytic task, while \textit{AI} adaptation entails reacting to feedback received from either the system or other participating agents. 
If \textit{both} types of agents have the capacity to learn, then the system can be considered co-adaptive. A recurrent limitation in this entire adaptation category is the difficulty in modeling human learning \refLFW{LFW1}.

\dimension{Adaptation Method}{adaptation}{How do the agents adapt?}
{Improve Task Performance, Improve Communication}

How agents adapt can be grouped into two comprehensive classes. \textit{Improving task performance} refers to all cases where the agents are getting better at solving the analytical task itself. 
For human agents, this could be reflected in increased efficiency due to becoming more familiar with the task, and for AI agents, it could involve fine-tuning the model parameters based on newly obtained information and feedback. 
The second way agents can adapt is by \textit{improving communication}. 
This encompasses everything related to conveying information to other agents. For example, an AI agent might include explanations when providing guidance to show its reasoning and thus communicate its suggestions better. 
Similarly, learning agent preferences, when it comes to things like modality or degree of guidance, help facilitate improved collaboration. 
While it might appear implausible to describe a multifaceted concept like learning with only two options, we postulate that they effectively cover the full scope of adaptation in human-AI interaction paradigms. 

\dimension{Information Learned}{adaptation}{What information do the agents learn?}
{Domain, Data, Task, Agent Goals, Agent Preferences}

This question supplements the aforementioned ones and details exactly what information the individual agents are obtaining to facilitate the adaptation process. 
Similarly, the distinction informs the extent or degree of learning that is happening, which is not reflected in the previous dimensions. 
Such a characterization is key for discerning between various adaptive frameworks and provides the capacity to discuss adaptation in more concrete terms. 
Continuing with our proposed terminology, \textit{data}, \textit{domain}, \textit{task} options can be seen as ways of improving task performance. While \textit{agent goals} and \textit{agent preference} reflect improvements in communication. This classification into two groups also has exceptions. For example, if the goal of the task is to align with human preferences, then this information pertains to both learning the task and improving communication. 

\begin{figure*}
  \centering
  \includegraphics[width=1\linewidth]{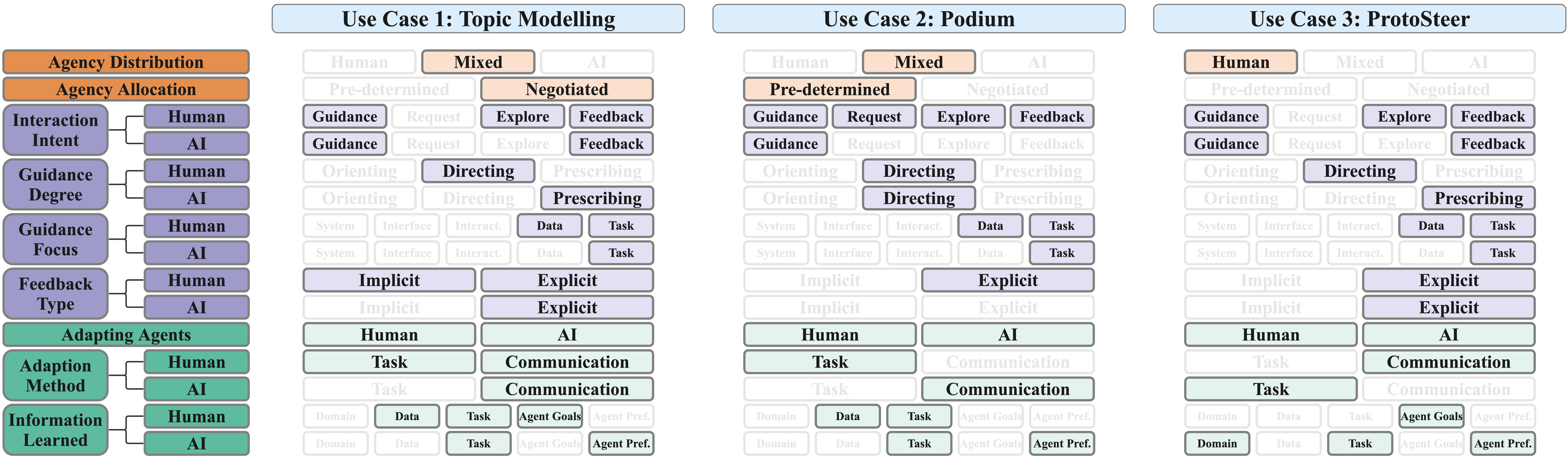}
  \vspace{.5em}
  \caption{Complete overview of the three use case systems positioned in our proposed design space. The interfaces of the three systems are depicted in \autoref{fig:usecases}. Each of the three systems exhibits some form of human-AI collaboration; however, these systems cover different areas of the proposed design space. Note that this overview analysis does not consider temporally changing dynamics, as pointed out in \refRO{RO2}. }
  \label{fig:studies}
  \vspace{-.5em}
\end{figure*}

\subsubsection*{Participant Comments and Interview Observations} 
\begin{leftbar}
Overall, this category evoked the greatest amount of agreement from the participants. 
Most were convinced by the number of dimensions and their increasing specificity. The main repeating themes were human learning and the role of feedback. 


\textbf{Human Learning ~|~}  Some participants remarked on the difficulty in describing learning and adaptation when dealing with human agents. Specifically, while it made sense to model AI learning as improving the task or communication, for the human side, this seemed too primitive. The various types of information learned are also less clearly separable for the human agents as they are often intertwined when reasoning about tasks. This human learning paradigm was decided to be out-of-scope for this conceptual model and thus accepted as a recognized limitation of the system that should be addressed in future iterations \defLFW{LFW1}. 

\textbf{Feedback ~|~}  Another aspect that was refined during the interviews was how feedback is tied to adaptation. Since, in most cases, learning is directly prompted by some type of feedback, there is an argument to include it under the adaptation term. However, since this connection is not bi-directional and not all feedback is concerned with learning, it was retained in the interaction category.  


\end{leftbar}

\begin{figure}
\centering
\begin{subfigure}{\columnwidth}
    \includegraphics[width=\textwidth]{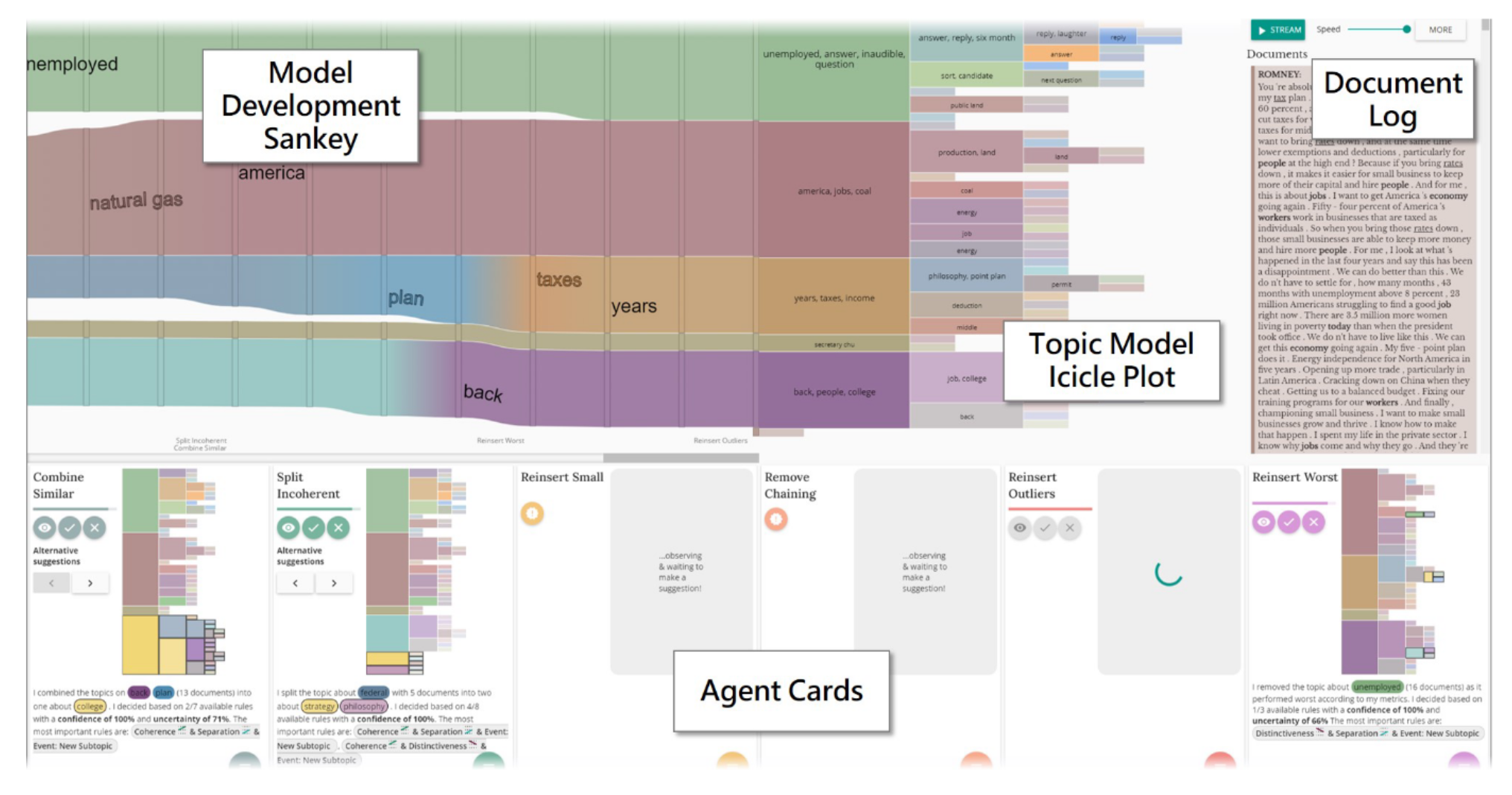}
    \caption{Topic Model Refinement~\cite{sperrle_learning_2021}.}
    \label{fig:topic-modeling}
\end{subfigure}

\begin{subfigure}{\columnwidth}
    \includegraphics[width=\textwidth]{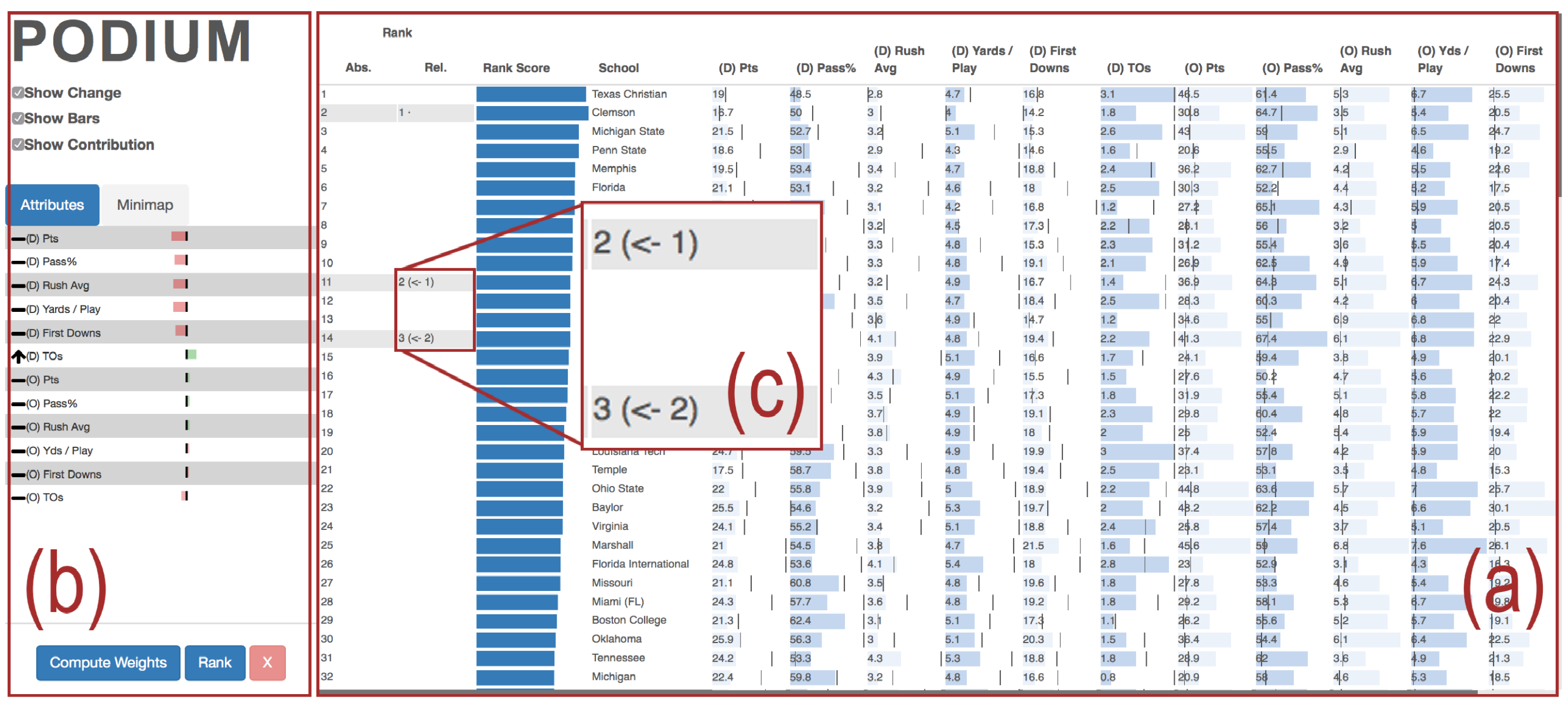}
    \caption{Podium~\cite{wall_podium_2018}.}
    \label{fig:podium}
\end{subfigure}

\begin{subfigure}{\columnwidth}
    \includegraphics[width=\textwidth]{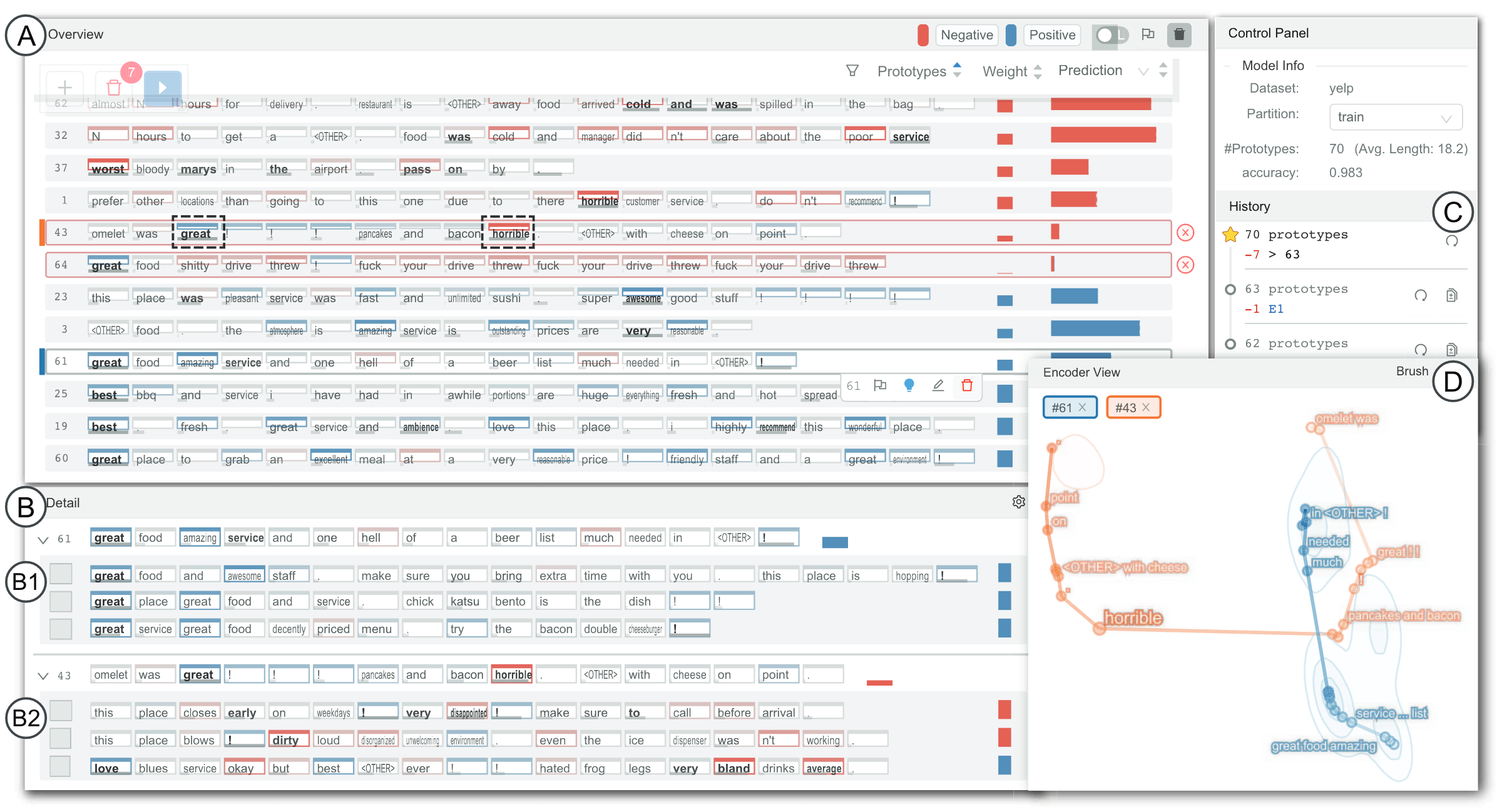}
    \caption{ProtoSteer~\cite{ming_protosteer_2020}.}
    \label{fig:protosteer}
\end{subfigure}
\caption{Overview of the visual interfaces of the three selected human-AI systems for the case studies described in \autoref{sec:usecases}.}
\vspace{-1.2em}
\label{fig:usecases}
\end{figure}


\section{Case Studies}\label{sec:usecases} 
Using three examples from the surveyed literature, we now demonstrate how our design space can be applied to characterize human-AI collaborative systems. 
The proposed conceptual model only holds weight if it can be applied freely to all systems within this paradigm. To exemplify our work, we selected papers that fulfilled two main criteria. Firstly, they provide the necessary level of detail regarding the three categories (i.e., agency, interaction, adaptation) to be able to be encoded effectively. Secondly, they incorporated sufficient complexity to highlight the capabilities of our conceptual model. For example, many popular VA tools were omitted since they could be easily modeled on just a few dimensions. 

For each use case, we give a general overview of the details of the system and then apply our conceptual model to describe the specific properties of the system. The main interfaces of all three systems are depicted in \autoref{fig:usecases}. A comparative view of all the systems discussed in this section and how they are positioned in our proposed design space can be seen in~\autoref{fig:studies} and is discussed in detail below.



\subsection{Co-Adaptive Guidance for Topic Model Refinement}

Sperrle et al.~\cite{sperrle_learning_2021} propose a system where six artificial agents assist analysts in refining topic models. A multi-objective optimization task that is subjective and context-dependent, such as topic model refinement, represents a prototypical task for human-AI collaboration. The two types of agents featured in the system are (1)~human users and (2)~six AI agents. While the AI agents contribute different refinement operations, their overall characteristics with respect to agency, interaction, and adaptation are identical, meaning no additional context can inferred from viewing them separately. The complete itemization of the system using our design space can be seen in~\autoref{fig:studies} and is discussed in detail below. The interface of the proposed system is depicted in \autoref{fig:topic-modeling}.

\noindent\textbf{Agency ~|~}  The system is defined as being \textit{mixed}-initiative, meaning that both types of agents can have agency. In the initial stages of the interaction, the human guides the process, and the AI agents simply collect data. However, after learning user preferences, the AI starts to automatically enact refinement operations. The agency allocation can be viewed as \textit{negotiated} as it is determined dynamically based on how the agents interact within the system. 

\noindent\textbf{Interaction ~|~}  The humans users' interaction intents are \textit{receiving guidance}, \textit{exploration}, and \textit{giving feedback}. The guidance is \textit{directing} and comes in the form of a ranked set of multiple small icicle plots and verbalizations recommending a certain operation and accentuating relevant info. Thus, the focus of the guidance is both task- and data-specific. Exploration is possible by exploring the data inputs, existing topics, and keywords. Finally, the human can provide \textit{explicit} feedback by consciously accepting or rejecting the agent's recommendations and \textit{implicitly} by repeatedly ignoring suggestions. The AI agent's interaction motives are also \textit{receive guidance} and \textit{provide feedback}. The inputs concerning preferences from the users is \textit{prescribing} guidance, and in response, the \textit{explicit} feedback is the learned optimization preferences for the current model state.

\noindent\textbf{Adaptation ~|~}  As a co-adaptive system \textit{both} human and AI agents are adapting during the collaboration. The human agent is improving in \textit{task performance} by learning about the \textit{data} and the \textit{task} through exploration and repeated topic refinement. It can be inferred that the human agent also obtains domain knowledge, but this isn't explicitly modeled by the described system. Similarly, through interacting with the different AI agents, the human is \textit{improving communication} as it learns AI \textit{goals}. On the other hand, the AI agent's adaptation is centered around \textit{improving communication} and learning context-dependent human \textit{user preferences}. Since the topic modeling task is by definition subjective and aligns directly with the human \textit{user preferences}, it can be argued that the AI is also adapting to the \textit{task}.

 \subsection{Podium} 

Podium~\cite{wall_podium_2018} is a visual analytics tool to assist users in ranking multi-variate data based on the perceived importance of the data points. In other words, Podium helps users quantify the importance of particular attributes to a decision. Once again, we will differentiate between the human user as one agent and the AI extracting relevant attributes and features as the other type of agent. The analysis for the system is shown in~\autoref{fig:studies}. The interface of the proposed system is depicted in \autoref{fig:podium}.

\noindent\textbf{Agency ~|~}  The distribution of agency is once again \textit{mixed}; however, the specifics differ from the previous example. In this case, mixed-initiative principles are employed through the automated use of complex mathematical models. The AI agent takes intuitive ranked inputs and automatically models these interactions to produce an attribute weight vector. Thereby, the agency is shared, and there is a better balance between human and AI efforts. The allocation of agency is \textit{pre-determined} since the assignments for when each type of agent is in control is embedded into the design of the system.  

\noindent\textbf{Interaction ~|~}  In Podium, the human user's interaction can be motivated by a range of intents, namely, \textit{receiving guidance}, \textit{requesting information}, \textit{exploration}, and \textit{providing feedback}. The guidance is primarily \textit{directing} as new rankings are generated based on user preferences. It could also be reasoned that the generated weight vector is a \textit{prescribing} type of guidance. The human user can explicitly request information in the form of new weight vectors and rankings. The exploration is facilitated by interacting with the data in the main view as well as various toggles for visual encoding. Finally, feedback is given by the human \textit{explicitly} through ordering the data points in the main table. The AI agent is more constrained in its intents and is limited to \textit{receiving guidance}. Furthermore, this received ranked set of data points can be classified as \textit{directing} guidance.


\noindent\textbf{Adaptation ~|~}  Both types of agents are adapting in this system. The human is primarily \textit{improving task performance} as they incrementally learn about the impact of specific features on decisions. The information learned is related to the \textit{data} and, consequently, the \textit{task}, which in this case is closely related to understanding the data properties. The AI agent's learning is focused on \textit{improving communication} and extracting \textit{user preference} information through repeated interaction with the human user. Similar to the topic refinement system described previously, the task itself is dependent on preferences, so the AI can be assumed to learn about the task, as well.


 \subsection{ProtoSteer} 
In the final use case, we use our design space to analyze and encode the ProtoSteer interface~\cite{ming_protosteer_2020}. The tool allows domain experts to steer a deep sequence model by making modifications to provided prototypes. Thus, the general analysis task for the agents involved is to improve the models' performance by either increasing accuracy or reducing the complexity of the prototypes. The collaboration consists of one human agent who is a domain expert and one AI agent that generates prototypes. The interface of the proposed system is depicted in \autoref{fig:protosteer}.

\noindent\textbf{Agency ~|~}  In this visual interface, the entire operation is coordinated by the \textit{human}; thus, there is no sharing of initiative. This also results in the allocation dimension being latent, as there is no need to distribute agency. However, describing this system could benefit from a higher granularity in the agency distribution spectrum \refRO{RO1} as it could be modeled as an AI-in-the-loop interface.

\noindent\textbf{Interaction ~|~}  For ProtoSteer, the human can interact to \textit{receive guidance}, \textit{explore}, or \textit{provide feedback}. The human is shown a set of prototypes with associated weights and significance measures as \textit{directing} guidance. This pertains to both the \textit{data} and the \textit{task}. The visual interface promotes exploration by integrating t-SNE projections to compare similarities between different prototypes. In response, the human can provide \textit{explicit} feedback to the AI by either adding, removing, or modifying the proposed prototypes. On the other end of the interaction paradigm, the AI \textit{receives guidance} and \textit{provides feedback}. Since the guidance is concrete modifications, it can be described as \textit{prescribing}. After processing the information, \textit{explicit} feedback is shown to the human by indicating neighborhood changes and showing how the prototype distribution changes as a result of the guidance.  

\noindent\textbf{Adaptation ~|~}  Adaptation is happening for both human and AI agents. In this collaboration, the human is simply \textit{improving communication} by learning about the AI \textit{agent's goals} and strategies. Since, by definition, ProtoSteer is designed to facilitate the transfer of human knowledge to the AI agent, it can be said that it does not get better at solving the task. However, this once again relates to the inherent complexity of modeling human learning \defLFW{LFW1}. As the AI agent is the subject of the analysis task, it therefore directly adapts to \textit{improve task performance}. The primary information learned concerns the \textit{domain} and \textit{task}. However, while the tool explicitly models the human expert's suggestions as task improvement, preferences are inherently still embedded in any human guidance. 

 \section{Discussion}

In this section, we reflect on the implications, limitations, future work, and opportunities in light of our design space and example case studies.
Overall, through the application of our design space to model the three systems, it became evident that to solve real-world analytical tasks, it is necessary to utilize intricate human-AI systems. One high-level insight provided by~\autoref{fig:studies} is how many of the single dimensions are utilized by each of these collaborative systems. Even an interface like ProtoSteer, which facilitates one analytical task - improving model prototypes - is outlined in multiple dimensions. It \change{can be postulated} that a human-AI system used to solve even more complex analytical tasks will feature dimensions from the full range of the design space (i.e., mixed-initiative, adaptation, etc). \change{Thus, evaluating the impact of individual dimensions on analytical task solving presents a future research opportunity.}




\subsection{Limitations and Future Work}

We proposed our conceptual model to capture the full extent of the human-AI collaboration \change{paradigm}. We do, however, acknowledge that there are limitations that should be addressed in future work.

\defLFW{LFW1}
\noindent\textbf{Cognitive Modelling} A key limitation is how we model learning for agents. In an attempt to uphold symmetry within the design space, an equivalent approach is used to depict learning in humans and AI. This, however, can be viewed as an oversimplification. To properly model human learning requires explicitly addressing cognitive processes. While this direction of interdisciplinary research might provide some insight, it exceeds the scope of this work.


\defLFW{LFW2} 
\noindent\textbf{Expressivity} 
Another limitation and avenue for future work is related to the granularity of the dimensions. For the properties modeled on a continuous spectrum, we limited ourselves to a select set of categorical values. By including more fine-grained options, we could increase the expressivity of the dimensions. For instance, we could supplement the agency distribution spectrum with intermediary options like human-in-the-loop and AI-in-the-loop. This additional expressivity comes at the cost of generalizability as classifying systems on a fine-grained spectrum is non-trivial. A clear set of specific definitions for all values would be essential to objectively apply such a design space. This exceeds the needs of this conceptual framework, but formulating a more practical framework might benefit from more elaborate and expressive dimensions.


\defLFW{LFW3} 
\noindent\textbf{Coding All Systems} In this paper, we explicitly model three collaborative systems collected in the literature review stage. In future work, we will look to encode the entire set of systems to construct a comprehensive corpus. Using this, we aim to characterize the current state of the field of human-AI collaboration and also get a comprehensive overview of the research gaps.

\subsection{Research Opportunities}

In the following, we identify three research opportunities as a call for action to the research community. 

\defRO{RO1} 
\noindent\textbf{Taxonomy of Agency} While the high-level aspects of interaction and adaptation have been addressed in literature at length, the topic of agency remains unexplored. At a general level, it is sufficient to consider only the distribution of agency, but there is considerably more nuance in this space. For instance, how to facilitate unbiased negotiation and how does the dynamics of shared agency affect things like communication. Also, what is the full scope of the trade-off between having control and automation? These questions will not only allow for more informed choices in future system design, but they will also allow practitioners to reason about systems in a more complete way. 

\defRO{RO2}  
\noindent\textbf{Modeling Temporal Dynamics} The current conceptual model assigns properties to systems by generalizing across all the separate interactions. However, this imposes a limit on how we can discuss collaborative systems that feature dynamic modes of interaction. Agency and adaptation are often correlated with the stage of interaction; thus, using our design space, we cannot distinguish between a system that learns only in the beginning and a system that is always improving. The challenge lies in categorizing the time dimension in a uniform way to generalize to different use cases.

\defRO{RO3}  
\noindent\textbf{Handling Complexity} In order to comprehensively study collaboration as we described in this paper, there is a need to engineer systems that allow testing different variations in the dimensions. To facilitate this, we will inherently need to design systems complex enough to model the entire design space. There is an opportunity to outsource this complexity by, for instance, creating a platform that handles the underlying details and frees up researchers' time to study the interplay between the dimensions themselves. 

 \section{Conclusion}

This paper presents a design space for describing human-AI collaborative systems. Through an iterative refinement process involving literature review and interviews, we established a set of dimensions to model this collaboration paradigm's various facets. The approach yields agency, interaction, and adaptation as the defining properties of these systems. Our contribution is exemplified through the application of this conceptual model to existing human-AI interfaces. The analysis of additional systems is available at \underline{\href{https://hai-ds.ivia.ch}{hai-ds.ivia.ch}}.

\bibliographystyle{eg-alpha-doi}
\bibliography{main}

\clearpage

\end{document}